\pdfoutput=1

\documentclass[11pt]{article}

\usepackage[preprint]{styles/acl}

\usepackage{times}
\usepackage{latexsym}

\usepackage[T1]{fontenc}

\usepackage[utf8]{inputenc}

\usepackage{microtype}

\usepackage{inconsolata}

\usepackage{graphicx}

\usepackage{enumitem}
\usepackage{booktabs}
\usepackage{array}
\usepackage{colortbl}

\usepackage{amsmath}
\usepackage{algorithm}
\usepackage{graphicx} 
\usepackage{algpseudocode}
\DeclareMathOperator*{\argmax}{arg\,max}

\title{Prompts as Auto-Optimized Training Hyperparameters: \\[0.5ex] Training Best-in-Class IR Models from Scratch with 10 Gold Labels}

\author{
 \textbf{Jasper Xian\textsuperscript{1}},\hspace{1mm}
 \textbf{Saron Samuel\textsuperscript{2}},\hspace{1mm}
 \textbf{Faraz Khoubsirat\textsuperscript{1}},\hspace{1mm}
 \textbf{Ronak Pradeep\textsuperscript{1}},
\\
 \textbf{Md Arafat Sultan\textsuperscript{3}},\hspace{1mm}
 \textbf{Radu Florian\textsuperscript{3}},\hspace{1mm}
 \textbf{Salim Roukos\textsuperscript{3}},\hspace{1mm}
 \textbf{Avirup Sil \textsuperscript{3}},
\\
 \textbf{Christopher Potts\textsuperscript{2}},\hspace{1mm}
 \textbf{Omar Khattab\textsuperscript{2}}
\\
\\
 \textsuperscript{1}University of Waterloo,\hspace{1mm}
 \textsuperscript{2}Stanford University,\hspace{1mm}
 \textsuperscript{3}IBM Research AI
}

\begin{document}
\maketitle
\begin{abstract}
We develop a method for training small-scale (under 100M parameter) neural information retrieval models with as few as 10 gold relevance labels. The method depends on generating synthetic queries for documents using a language model (LM), and the key step is that we automatically optimize the LM prompt that is used to generate these queries based on training quality. In experiments with the BIRCO benchmark, we find that models trained with our method outperform RankZephyr and are competitive with RankLLama, both of which are 7B parameter models trained on over 100K labels. These findings point to the power of automatic prompt optimization for synthetic dataset generation.
\end{abstract}

\section{Introduction}

The past few years have witnessed massive improvements in information retrieval (IR) quality, thanks to many ways of applying and supervising pretrained language models (LMs) for IR. However, almost all current IR models, especially those known to generalize well to new long-tail domains, are trained on hundreds of thousands or even millions of queries and relevance judgments. From a scientific perspective, it is unclear if this magnitude of data is necessary for optimizing LMs for tasks like IR. At the same time, from an engineering standpoint, it remains unclear how to best train IR models for extremely long-tail domains or languages for which labeled IR data is scarce.

Motivated by these questions, we study the difficult problem of training an IR system from scratch, given nothing but a text corpus of passages $C$ and as few as 10 relevance judgments. To study this problem with limited confounders, we train only pretrained LMs that have under 100M parameters and 
have not been explicitly trained on labeled IR datasets like MS MARCO~\cite{msmarco} or undergone any similar supervision to the best of our knowledge.

An increasingly common paradigm to tackle the lack of IR data in a given domain is to use LMs to synthesize hypothetical search queries $q$ that are derived from passages $p$ in a corpus $C$. For example, Promptagator~\cite{dai2023promptagator} and UDAPDR~\cite{saad-falcon2023udapdr} use LMs like GPT-3 and Flan-T5~\cite{flan-t5} to generate queries. Each  query--passage pair $\langle q, p \rangle$ becomes a relevant training item, negative passages are sampled from $C$, and the resulting dataset is used to train an IR model.

However, such work either uses static prompts for LMs (Promptagator) or constructs prompts in automatic but static ways (UDAPDR). As a result, the LM-based data generation process receives no feedback from the IR model trained.
This is a substantial limitation; many IR tasks, like those in BIRCO~\cite{wang2024birco} such as ArguAna~\cite{arguana}, 
have nuanced notions of relevance, so simply relying on the priors of an LM for synthesizing queries may not suffice.

\begin{figure}
    \centering
  \includegraphics[width=0.98\columnwidth]{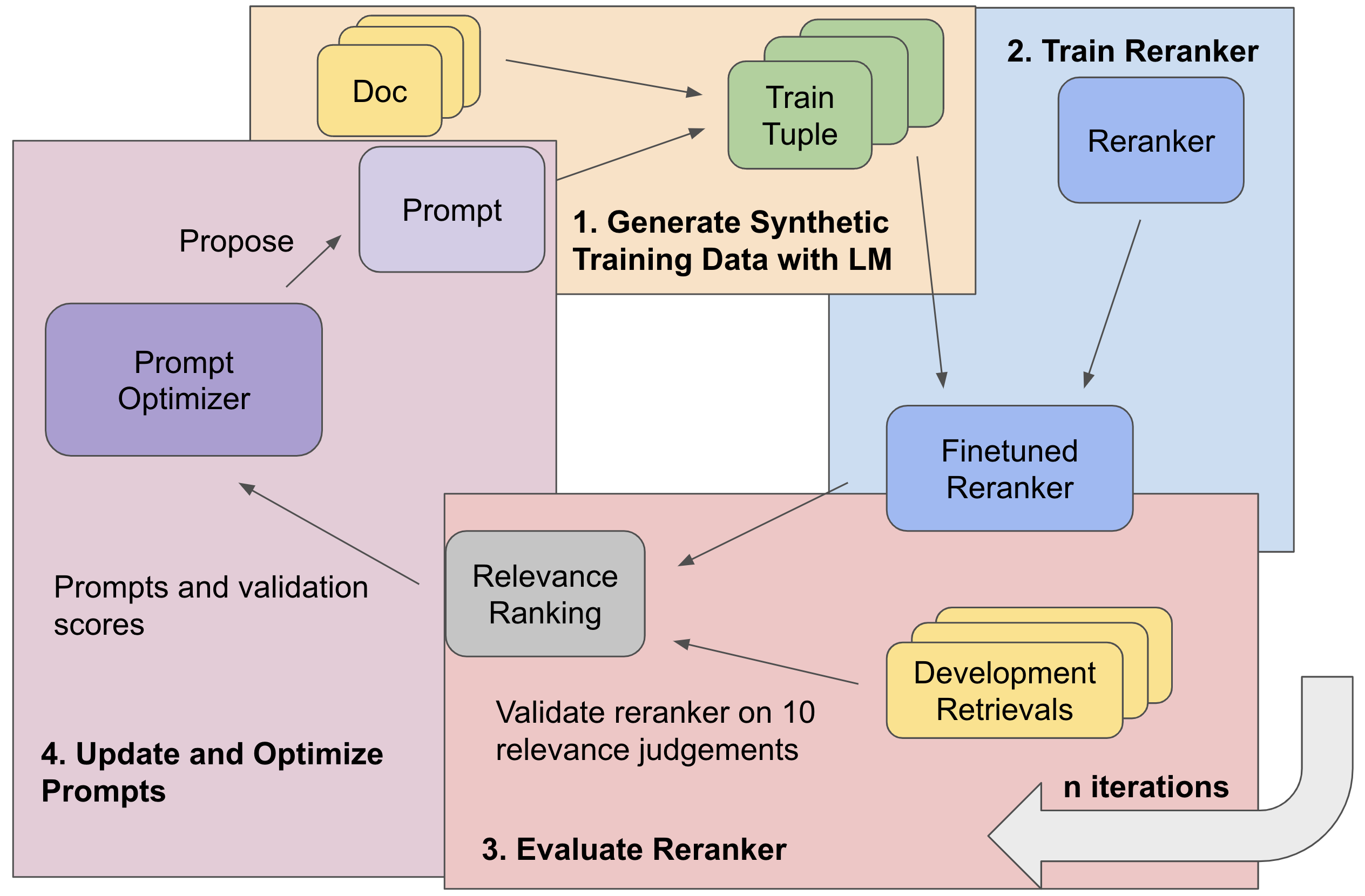}
  \caption{An overview of the PATH pipeline for training a reranker with synthetic queries. A user only needs to input a prompt with the task description and as few as 10 relevance judgements to achieve strong results.}
  \label{fig:pipeline}
  \vspace{-4mm}
\end{figure}

To overcome this, we propose \textbf{PATH}, for \textit{Prompts as Auto-optimized Training Hyperparameters}, to optimize the prompt used to generate synthetic queries. Figure~\ref{fig:pipeline} provides an overview of PATH. Steps~1--3 represent a typical pipeline for training IR models on synthetic data generated by an LM. Step~4 is the crucial addition: the prompt to the LM is updated using feedback from the reranker evaluation. 
In this paper, we adopt the simple strategy of having the LM generate candidate modifications of our initial instruction and choosing the one that ultimately leads to the best reranker.

We express PATH in the DSPy programming model~\cite{khattab2024dspy}, which allows us to treat the prompt responsible for synthesizing the queries as a parameter to learn. For optimization, DSPy requires a scalar metric. For the first time, we propose the metric of using the generated outputs of the prompt, i.e., the synthetic queries, to train an IR model and then returning the average quality of the resulting IR model directly as the score. This scoring can use as few as 10 gold labels.

We evaluate this idea using DeBERTa~\cite{he2021deberta} and MiniLM~\cite{wang2020minilm} on the BIRCO benchmark for difficult and non-traditional IR tasks.
We leverage gpt-3.5-turbo for query generation and prompt optimization. We find that applying PATH with 10 positive labels performs very competitively. 
In particular, averaged in NDCG@10 across tasks and LMs, it outperforms BM25 by 6.0 points, fine-tuned LMs on the 10 positive labels by 6.5 points, and hand-prompting GPT-3.5 for synthetic query generation by 4.5 percentage points.
Moreover, our approach performs roughly at the same level of quality as directly training on all available training triples for each task and are competitive with the best available off-the-shelf cross-encoders like MonoT5 and RankLLaMA, which are orders of magnitude larger in both parameter count and training set size.

\section{Preliminaries}

Given a large corpus of documents $\mathcal{D}=\{d_1, d_2, \ldots, d_n\}$ and a query $q$, a \emph{retriever} is a model that can find an ordered set $\mathcal{S}$ of $k \ll |\mathcal{D}|$ documents that are most \textit{relevant} to $q$, as measured using some metric(s) of relevance such as nDCG or recall.
We focus on the downstream task of \textit{reranking} the documents in $\mathcal{S}$ into a more accurate ordering. In particular, we are interested in training Transformer encoder models as \textit{point-wise rerankers}, i.e.\ a model $\mathcal{R}$ that can take the query and each document $d_i \in \mathcal{S}$ in isolation, $\mathcal{R}(q, d_i)$, and output a scalar score that can be used for re-ordering $\mathcal{S}$ to achieve higher relevance.

When only few labeled query-document pairs are available, synthetic data generation can augment the training dataset of a reranker~\cite{dai2023promptagator,inpars}. %
Canonically, this often involves sampling a subset of the documents in $\mathcal{D}$ randomly and, with some prompt template $\mathcal{P}$, asking an LM to generate a relevant query $q_s$ for each $d=d^+$ in our sample. Each synthetic query is then paired with its source document to create a $(q_s, d^+)$ positive tuple, which is then augmented by producing a set of $m$ hard negatives $(d^-_1, d^-_2, \ldots, d^-_m)$, sampled from the non-positive top results of an existing retriever. This process outputs a set of tuples $(q_s, d^+, d^-_1, d^-_2, \ldots, d^-_m)$ and uses them to train the reranker.

\begin{algorithm}[tp]
\begin{minipage}{\columnwidth}
\fontsize{8pt}{9pt}\selectfont
\caption{PATH for training a reranker with a small number, $N$, of relevance judgments.}
\label{alg:copro_algorithm}
\begin{algorithmic}[1]
\State \textbf{Input:} Large Autoregressive Language Model $\mathbf{LM}$
\State \textbf{Input:} Small Pretrained Encoder Model $\mathbf{Enc}$
\State \textbf{Input:} Document Corpus $\mathcal{D}$
\State \textbf{Input:} Number of Trials $M$, number of negatives $m$
\State \textbf{Input:} Relevance Judgments $\mathcal{J} = \{ \langle q_i, d_i, r_i \rangle : i \in [N] \}$
\State \textbf{Input:} Initial Prompt Instructions $\mathcal{I}_0$ for query generation \\

\Function{TrainReranker}{%
Prompt Template $P$}
    \State Sample a random subset $\mathcal{D}' \subseteq \mathcal{D}$, where $|\mathcal{D}'| = 1000$
    \State Triplets $\mathcal{T} \gets \{\}$
    \For{$d^+ \in \mathcal{D}$}
        \State Synthetic Query $q \gets \mathbf{LM}$.generate($P$.format($d$))
        \State Negatives $d^-_1, \ldots, d^-_m \gets$ SampleNegatives($\mathcal{D}$, $q$, $d^+$) %
        \State Extend $\mathcal{T}$ with $\{ \langle q, d^+, d^-_j \rangle : j \in [m] \}$
    \EndFor
    \State Reranker $\mathcal{R} \gets \mathbf{Enc}$.trainOnTriplets($\mathcal{T}$)
    \State \Return $\mathcal{R}$
\EndFunction
\\
\Function{AvgNDCG}{$\mathcal{R}$, $\mathcal{J}$}
    \State T = \{ $\mathcal{R}$.rerank($q_i$,  BM25.retrieve($q_i$)) : $q_i \in \mathcal{J}$.queries() \}
    \State \Return $\frac{1}{N} \sum_{i=1}^{N}$ NDCG(T$_i$, $\mathcal{R}$)
\EndFunction
\\
\State Initialize Attempts List $\mathcal{A} \gets \{\}$ 

\While{i in [$M$]}
    \State $\mathcal{P}_i \gets \textsc{proposeNewPrompt}(\mathcal{I}_0, \mathcal{A})$ \Comment{See Sec~\ref{sec:main} for how}
	\State $\mathcal{R}_i \gets \textsc{TrainReranker}(\mathcal{P}_i)$
	\State Validation Score $s \gets \text{AvgNDCG}(\mathcal{R}_i, \mathcal{J})$
	
	\State Extend $\mathcal{A}$ with $(s, \mathcal{R}_i, \mathcal{P}_i)$
\EndWhile
\\
\State Let the selected reranker be the best-scoring $\mathcal{R}_i$
\end{algorithmic}
\end{minipage}
\end{algorithm}

\section{PATH: Training Rerankers With Optimized Data-Generation Prompts}
\label{sec:main}

Algorithm~\ref{alg:copro_algorithm} describes our method for training rerankers using a very small number of task-specific relevance labels. Our goal is to train reranker $\hat{\mathcal{R}}$ that would have high quality on the (unseen) underlying distribution of $\mathcal{J}$. Unfortunately, simply training on the labels in (the very few labels in) $\mathcal{J}$ will result in drastic overfitting (Sec~\ref{sec:evaluation}). Our algorithm resolves that as follows.

We require access to (i) a large, autoregressive $\mathbf{LM}$ for prompting and (ii) a small pretrained encoder model $\mathbf{Enc}$ that we will finetune as a reranker. The algorithm takes as input two task-specific human inputs. (1) A small set, like 10 labels, of relevance judgments $\mathcal{J}$, each indicating a realistic query and a document that is assigned some relevance grade like 0 (irrelevant) or 3 (perfectly relevant).\footnote{For simplicity, we assume the provided judgments are all positive, i.e. $r \geq 1$. These relevance scores enable relevance metrics like NDCG@10 to assign a score to the ranking $\mathcal{S}$ established by a reranker on a given query, relative to the ideal ranking which places the documents assigned the largest relevance grades first.} (2) An initial instruction $\mathcal{I}_0$, which may be task-aware, for the $\mathbf{LM}$ generating queries, e.g. ``Given a passage, return a question a user may ask that is answered by this passage''.

The core of the algorithm is \textsc{TrainReranker}, a pipeline for generating synthetic queries (Line 11) and building triples (Lines 12--13) to train a point-wise Transformer encoder as a reranker $\mathcal{R}$ (Line 15). Crucially, the training of $\mathcal{R}$ depends on a prompt template $\mathcal{P}$, responsible for instructing the $\mathbf{LM}$ on the nature of the queries it must synthesize. Our central contribution is that we seek to automate the construction of a prompt template $\hat{\mathcal{P}}$ that maximizes the quality of the resulting $\mathcal{R}$.

\vspace{-4pt}

{\small
\begin{equation*}
  \hat{\mathcal{P}} =
  \argmax_{\mathcal{P}} 
  \, \textsc{AvgNDCG}(\textsc{TrainReranker}(\mathcal{P}), \mathcal{J})
  \label{eq:prompt_components_new}
\end{equation*}
}

This is essentially the problem of automatic hyperparameter optimization: automatically tuning $\mathcal{P}$ so that training $\mathcal{R}$ with gradient descent achieves a high score. However, we uniquely have a \textit{string template} as a hyperparameter. Optimizing a string prompt is a difficult problem that has been studied extensively in the past few years. We do not propose a new prompt optimization algorithm nor do we claim that a specific optimizer works best. Instead, we show that very simple choices about prompt optimization are sufficient to find a prompt $\hat{\mathcal{P}}$ that allows us to produce very high quality $\hat{\mathcal{R}}$.

To this end, we express Algorithm~\ref{alg:copro_algorithm} in the DSPy framework~\cite{khattab2024dspy}, which provides a suite of tools to algorithmically optimize LM prompts and weights in the context of larger programs. These tools can be thought of as instantiating the abstract \textsc{ProposeNewPrompt} in Algorithm~\ref{alg:copro_algorithm}. Concretely, we express \textsc{TrainReranker} as a DSPy program with one Chain-of-Thought ~\cite{wei2022chain} layer, which takes in each sampled passage and outputs a syntheized query.
We use one of DSPy's simplest prompt optimizers, \texttt{CA-OPRO},\footnote{This is an extension of the OPRO algorithm~\cite{yang2023large}, which generalizes it via Coordinate Ascent (CA) so it can apply to \textit{multi-prompt} programs as well as to scenarios in which quality is measured via a reward metric, like \textsc{AvgNDCG}, rather than a pre-defined correct output.} which iteratively refines the initial instruction $\mathcal{I}_0$ using suggestions by the $\mathbf{LM}$.

In our primary experiments, we set \texttt{CA-OPRO}'s $\texttt{depth}=1$, which reduces \textsc{ProposeNewPrompt} (Line 26) to simply feeding the $\mathbf{LM}$ our initial instruction $\mathcal{I}_0$ and asking it to produce a new proposed instruction that leads to higher accuracy. In this simplest instantiation of Lines 24--30, our algorithm simply tries $M=10$ different (automatic) prompt variants, executing \textsc{TrainReranker} to synthesize queries and train a new reranker each time. This process is quick since we use very small encoders $\mathbf{Enc}$, trained on small synthetic sets. The reranker $\hat{\mathcal{R}}$ that scores highest on \textsc{AvgNDCG} is then selected for deployment and returned for held-out evaluation (Sec~\ref{sec:evaluation}). In the appendix, we report the before-and-after prompts (Table~\ref{tab:appendix-prompts-demo}). 

Many other optimizer choices are possible. For example, in the appendix (Table~\ref{table:full_devset}), we report successful application of \texttt{CA-OPRO}'s $\texttt{depth}=2$ (Figure~\ref{fig:meta-prompts}), which allows richer feedback to flow back to the $\mathbf{LM}$ proposing instructions. %
In particular, \textsc{ProposeNewPrompt} (Line 26) will now see the prompts it generated earlier and how well they performed on \textsc{AvgNDCG}, essentially creating momentum in the right prompting direction. Other optimizers in DSPy work by crafting examples (e.g., of queries that have been effective) instead of instructions or even by updating the weights of $\mathbf{LM}$. These all suggest straightforward extensions of our method but we leave them for future work.

\section{Evaluation}
\label{sec:evaluation}
\begin{table}
\small
\renewcommand{\arraystretch}{1.2} %
\centering
\setlength{\tabcolsep}{2.5pt}
\scalebox{0.85}{
    \begin{tabular}{lccccc|c}
         \toprule
          & ArguA & CTrial & DMAE & Relic & WTB & AVG\\
         \hline
         (0) BM25 & 35.0 & 9.9 & 52.6 & 10.1 & 16.5 & 24.8 \\ [0.5ex]
         \hline
         \multicolumn{5}{l}{(1) \textit{Training directly using $N=10$ judgments}} \\
         \hline
         DeBERTA-v3 (86M) & 34.8 & 7.3 & 45.5 & 14.1 & 16.3 & 23.6 \\
         MiniLM-L12 (33M) & 44.2 & 6.0 & 46.0 & 11.8 & 17.8 & 25.2 \\ [0.5ex]
         \hline
         \multicolumn{5}{l}{(2) \textit{Manual Prompting for Synthetic Queries}} \\
         \hline
         DeBERTA-v3 (86M) & 41.6 & 14.7 & 48.0 & 12.2 & 16.9 & 26.7 \\
         MiniLM-L12 (33M) & 38.0 & 13.2 & 50.5 & 9.2 & 19.3 & 26.0 \\ [0.5ex]
         \hline
         \multicolumn{5}{l}{(3) \textit{Unoptimized DSPy Synthetic Queries}} \\
         \hline
         DeBERTA-v3 (86M) & 40.5 & 15.5 & 55.5 & 14.0 & 18.0 & 28.7 \\
         MiniLM-L12 (33M) & 33.9 & 15.2 & 54.6 & 11.5 & 19.7 & 27.0 \\ [0.5ex]
         \hline
         \multicolumn{7}{l}{(4) \textit{\textbf{PATH}: Optimized with DSPy \texttt{CA-OPRO} via $N=10$ judgments}} \\
         \hline
         \rowcolor{gray!20}
         DeBERTA-v3 (86M) & \underline{\textbf{49.7}} & 14.3 & \textbf{57.1} & \underline{\textbf{15.3}} & 23.4 & \textbf{32.0} \\
         \rowcolor{gray!20}
         MiniLM-L12 (33M) & 40.6 & \textbf{14.9} & 55.5 & 12.4 & \textbf{25.3} & 29.7 \\ [0.5ex]
         \hline
         \multicolumn{7}{l}{\color{gray} (5) \textit{Reference Rerankers, trained on massive data like MS MARCO}} \\
         \hline
        {\color{gray} monoT5 (220M)} & {\color{gray} 25.7} & {\color{gray} 15.3} & {\color{gray} 53.6} & {\color{gray} 12.7} & {\color{gray} 18.3} & {\color{gray} 25.1} \\
        {\color{gray} monoT5 (3B)} & {\color{gray} 39.8} & {\color{gray} \underline{17.6}} & {\color{gray} 61.2} & {\color{gray} 11.2} & {\color{gray} 30.8} & {\color{gray} 32.1} \\
        {\color{gray} RankZephyr (7B)} & {\color{gray} 35.2} & {\color{gray} 15.7} & {\color{gray} 65.9} & {\color{gray} 10.4} & {\color{gray} 29.4} & {\color{gray} 31.3} \\
        {\color{gray} RankLlama (7B)} & {\color{gray} 41.6} & {\color{gray} 13.7} & {\color{gray} \underline{66.8}} & {\color{gray} 15.0} & {\color{gray} 35.7} & {\color{gray} \underline{34.6}} \\
         \bottomrule
    \end{tabular}
}
\caption{nDCG@10 on BIRCO with CA-OPRO, other baselines, and various rerankers. All rerankers are pointwise except RankZephyr which is listwise. We use a window size of 20 and a step size of 10 for RankZephyr. In any setting that we use 10 positive labels, we average nDCG@10 over 7 different samples and runs. The best results overall are underlined, and the best results with DeBERTA and MiniLM are bolded.} 
\label{table:main}
\end{table}

We use the BIRCO benchmark for information retrieval, a collection of five complex QA tasks, each with a development dataset and a test dataset. All final results shown are on the full held-out test set. For Algorithm~\ref{alg:copro_algorithm}, we sample $|\mathcal{J}|$ positive relevance judgments, which in our primary experiments are $N=10$, for prompt optimization.  We use \texttt{gpt-3.5-turbo} as the $\mathbf{LM}$. Each passage is used to generate one synthetic query, which is used to mine $m=19$ hard negatives randomly sampled from the top 20 to 100 hits retrieved by BM25. %

For our $\mathbf{Enc}$ encoder models, we choose \texttt{MiniLM-L12-H384-uncased} (33M backbone parameters) and \texttt{DeBERTA-v3-base} (86M backbone parameters), both known for strong performance relative to parameter size. %
We train each on the synthesized triples (Line 16) using LCE cross-entropy loss~\cite{gao2021lce} over 2 epochs, validating \textsc{AvgNDCG} on $\mathcal{J}$ every half-epoch.
As our goal is to evaluate rerankers trained \textit{without} large collections like MS MARCO, we use BM25 as our initial retriever. All methods rerank the top-50 BM25 document retrievals.

We consider several baselines. Baseline (1) evaluates the idea of using the $N=10$ available positive judgments to create training triplets equivalently to Lines 10--16 of Algorithm~\ref{alg:copro_algorithm} \textit{but using only the real positive queries from $\mathcal{J}$} instead of generating any synthetic queries (Line 12).\footnote{In all settings in which we train directly on top of judgments, we train for 2000 steps, mirroring the number of training tuples seen during each iteration of training within PATH. In cases where the development dataset is very small (i.e., 2000 training steps is greater than 10 epochs of training), we limit training to 10 epochs.} Baseline (2) shows a more standard approach for training in low-data regimes, which is to manually prompt our $\mathbf{LM}$ to produce synthetic queries, i.e. the \textsc{TrainReranker} algorithm invoked with a manual prompt template $\mathcal{P}$. Baseline (3) shows a simple variant of that, which uses an (unoptimized) version of the \textsc{TrainReranker} algorithm expressed with a DSPy Chain-Of-Thought pipeline, but receiving no feedback from the IR model. Finally, we also include Baseline (5) which is a collection of large, popular reference rerankers that we evaluate on BIRCO. These models are given access to much more IR data for training, so their role here is to serve as reference points for high-quality out-of-the-box performance.

\section{Results \& Discussion}

Table~\ref{table:main} reports our primary results. Methods (1) and (4) involve sampling $N=10$ judgments, so we re-sample and re-run for a total of 7 times and report the average score in each cell. 
With only 10 labels, PATH leads to training a reranker that performs an average of 4.5 nDCG@10 points better than manually-written prompts and 3.0 points better than DSPy unoptimized queries across all datasets.
The biggest improvement came in ArguAna~\cite{arguana} and DORIS-MAE~\cite{wang2023dorismae}, with improvements of almost 10.0 points each on DeBERTA.
We also see that, with only 10 labeled relevance judgements, it is far better to use them with PATH as opposed to directly training with the labels.
Training directly with labels yields worse perfomance on average (by 6.5 points) and on each dataset split, particularly Clinical-Trial~\cite{clinical-trial}.

Our small rerankers trained with PATH-generated tuples are also competitive with current state-of-the-art LM rerankers trained on large datasets such as MS MARCO~\cite{msmarco}. For instance, with PATH and 10 labels, a finetuned DeBERTA outperforms state-of-the-art models on ArguAna and Relic~\cite{thai-etal-2022-relic}, which are relatively complicated QA tasks. 
Notably, with DeBERTA we outperform RankZephyr~\cite{pradeep2023rankzephyr} and RankLLama~\cite{ma2023finetuning}, which are 7B models trained on MS MARCO, on ArguAna by 14.5 and 8.1 points respectively.
We also outperform 3B models monoT5-3B~\cite{nogueira-etal-2020-document} as well as UPR~\cite{sachan2022improving}, whose reranker uses T0\_3B~\cite{sanh2021multitask},  by similar margins. 
In contrast, the stable of billion-parameter rerankers perform well on DORIS MAE and WhatsThatBook~\cite{wtb}, which are comparatively more straightforward relevance tasks akin to the datasets they were trained on.

\section{Limitations}

This work uses one possible set of many potential hyperparameters that may affect the performance of PATH. 
We only ran full experiments with one initial, human-written prompt per task, and it is unclear how changing that will affect downstream performance. 
We also use a fixed learning rate (5e-5) and warmup ratio (0.1), amongst other hyperparameters, across each experiment. 
These are examples of potential hyperparameters that can be optimized in future work.
We also define an arbitrary floor for ``positive'' relevance in the DORIS-MAE dataset~\cite{wang2023dorismae}, as DORIS-MAE has multi-level floating point relevance.  This floor has been manually tuned for Baseline (7) in Table~\ref{table:full_devset}, but not for PATH. Different choices for defining positive relevance in DORIS-MAE may yield differing results.
We used Tesla-V100s and Titan-V GPUs for our experiments. Our work is done with small, sub-100M parameter models, and we encourage future work to extrapolate to larger, billion-parameter models, which may achieve even higher quality.

\section{Acknowledgements}
We thank Jimmy Lin and Martin Franz for their valuable guidance and feedback.

\bibliography{custom}

\clearpage
\newpage
\appendix

\section{PATH Results with Full Development Set}
\label{sec:full-devset}

\begin{table}
\small
\renewcommand{\arraystretch}{1.2} %
\centering
\setlength{\tabcolsep}{2.5pt}
\scalebox{0.8}{
    \begin{tabular}{lccccc|c}
         \toprule
          & ArguA & CTrial & DMAE & Relic & WTB & AVG\\
         \hline
          \multicolumn{2}{l}{(6) \textit{Base Models}} \\
         \hline
         DeBERTA-v3-base (86M) & 5.9 & 7.4 & 41.8 & 2.4 & 3.9 & 12.3 \\
         MiniLM-L12 (33M) & 22.7 & 9.9 & 50.1 & 8.7 & 10.6 & 20.4 \\ [0.5ex]
         \hline
         \multicolumn{6}{l}{(7) \textit{Training directly using all relevance judgements}} \\
         \hline
         DeBERTA-v3-base (86M) & 60.0 & 8.4 & 49.0 & 22.0 & 19.8 & 31.8 \\
         MiniLM-L12 (33M) & 58.5 & 6.7 & 62.4 & 16.6 & 20.0 & 32.8 \\ [0.5ex]
         \hline
         \multicolumn{7}{l}{(8) \textit{\textbf{PATH}: Optimized with DSPy \texttt{CA-OPRO} at depth=1 via all judgments}} \\
         \hline
         DeBERTA-v3-base (86M) & 48.2 & 13.9 & 58.9 & 16.5 & 23.9 & 32.3 \\
         MiniLM-L12 (33M) & 44.3 & 14.4 & 58.0 & 13.1 & 26.3 & 31.2 \\ [0.5ex]
         \hline
         \multicolumn{7}{l}{(9) \textit{\textbf{PATH}: Optimized with DSPy \texttt{CA-OPRO} at depth=2 via all judgments}} \\
         \hline
         DeBERTA-v3-base (86M) & 48.8 & 13.7 & 61.7 & 16.0 & 23.9 & 32.8 \\
         MiniLM-L12 (33M) & 44.6 & 14.4 & 60.1 & 13.3 & 32.0 & 32.9 \\
         \bottomrule
    \end{tabular}
}
\caption{nDCG@10 on BIRCO with PATH given access to the full development sets.} 
\label{table:full_devset}
\vspace{-4mm}
\end{table} %
Table~\ref{table:full_devset} shows the performance of PATH given at various depths given access to the entire BIRCO development set.
On average, PATH at $\texttt{depth}=2$ performs 0.5 points better than directly training with the full development set. 

\section{Meta Prompting with \texttt{CA-OPRO}}

\definecolor{burntorange}{rgb}{0.8, 0.33, 0.0}
\definecolor{ceruleanblue}{rgb}{0.16, 0.32, 0.75}
\definecolor{forestgreen}{rgb}{0.13, 0.55, 0.13}

\begin{figure}
  \includegraphics[width=\columnwidth]{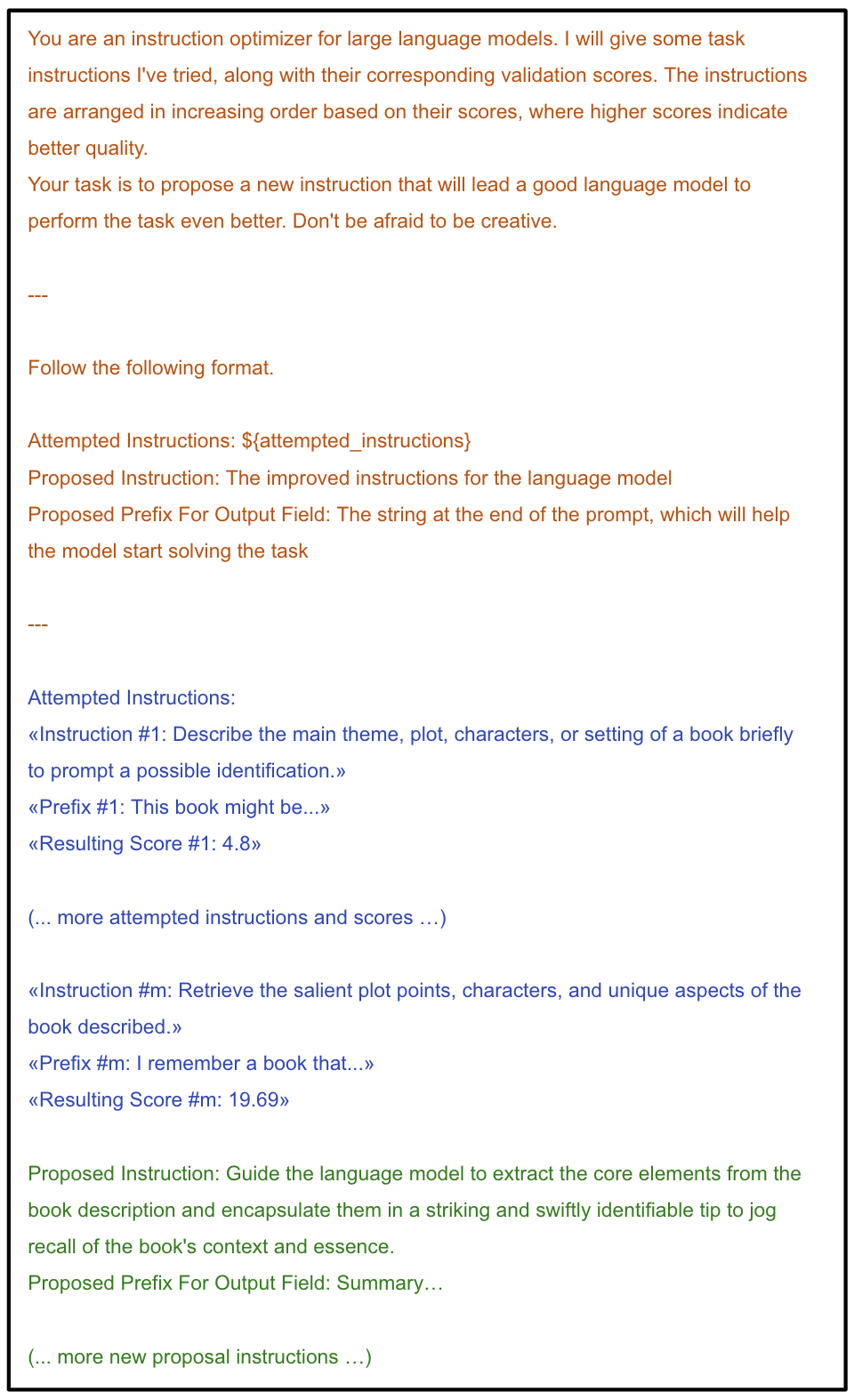}
  \caption{An example of \texttt{CA-OPRO}'s meta-prompts for prompt optimization. \textcolor{burntorange}{Orange} text represents the meta-prompt, \textcolor{ceruleanblue}{blue} text represents attempted trial instructions, and \textcolor{forestgreen}{green} text represents \texttt{CA-OPRO}'s new proposed instructions.}
  \label{fig:meta-prompts}
  \vspace{-4mm}
\end{figure}

Figure~\ref{fig:meta-prompts} shows an example of the meta-prompting strategy used by \texttt{CA-OPRO}. 
The proposed optimal instructions are sent back through \textsc{TrainReranker} and evaluated again by \textsc{AvgNDCG}.
This interaction occurs before each depth level > 1, thereby repeating until ideally converging on more optimal instructions.

\section{Analysis of PATH Prompts}

\begin{table*}[t]
    \centering
    \small
    \scalebox{0.9}{
    \begin{tabular}{ | p{3em} | p{15em}| p{15em} | p{10em} | p{5em} | }
    \hline
    Task & Initial Manual Prompt & Final PATH-optimized Prompt & Final PATH-optimized Suffix & nDCG@10 Improvement \\
    \hline
    ArguA  & Given a passage with an argument, please return the best counterargument that refutes the input passage. The counterargument should be a few sentences long. Only return the counterargument; do not reason or explain. &  Generate a succinct counterargument that refutes the input passage. & Counterargument: & 7.2 \\
    \hline
    CTrial  & Given a passage with the description of a clinical trial, return the a patient record that would match that of the required subjects for the input clinical trial. Please describe this patient record in a few sentences. & Instruction \#11: Starting with the clinical trial description, create a comprehensive patient record containing demographic information, medical history, current medications, and any other pertinent details professionally arranged to capture the essence of the trial's requirements. & Comprehensive Patient Record: & -1.0 \\
    \hline
    DMAE  & Given a passage consisting of an abstract of a computer science paper, please return a complex, multiple-sentence research question that is best answered by the input abstract. Only return the question; do not reason or explain. &  Given an abstract of a computer science paper, construct a multi-dimensional research question that delves deeply into the topic and explores new dimensions beyond the provided information. Integrate analysis of the main goals and contributions along with potential areas for further study. & Further Inquiry: & 13.7 \\
    \hline
    Relic  & Given a literary quotation, return an excerpt of text that is most likely to include the input quotation within it. The output excerpt should include the token [masked sentence(s)] replacing the input quotation, as well as a few sentences before and after the token. Only return the excerpt; do not reason or explain. & Review several paragraphs surrounding the given literary quotation and intelligently formulate an excerpt that seamlessly integrates the quote. Ensure the generated text captures the essence and context of the input quotation authentically. & Literary Excerpt: & 3.8 \\
    \hline
    WTB  & Given the description of a book, please return a tip-of-the-tongue description that someone might use in order to try and identify the book the input describes. The output should be in first-person. Only return the tip-of-the-tongue description; do not reason or explain. Follow the following format. &  Consider the key plots, characters, and themes of the book to generate a memorable and concise tip-of-the-tongue description reflecting its essence. Avoid using explicit titles or characters in the output to encourage creative thinking in forming the tip-of-the-tongue description. & I'm thinking of a book that... & 7.0 \\
    \hline
    \end{tabular}
    }
    \caption{A comparison of initial manual prompts and final, PATH-optimized prompts.}
    \label{tab:appendix-prompts-demo}
\end{table*}

We visualize the effects of PATH in Table~\ref{tab:appendix-prompts-demo}. 
The initial manual prompts were used in Baselines (2) and (3) in Table~\ref{table:main}, and the final PATH prompts are drawn from training DeBERTA in Baseline (9) in Table~\ref{table:full_devset}.

\end{document}